\pdfoutput=1
\documentclass[]{spie}  %>>> use for US letter paper
\usepackage[]{graphicx}
\usepackage[]{aas_macros}

\title{Investigating reciprocity failure in 1.7-micron cut-off HgCdTe detectors} 

\author{
M.~Schubnell\supit{a},
T.~ Biesiadzinski\supit{a},
W.~Lorenzon\supit{a},
R.~Newman\supit{a},
G.~Tarl\'e\supit{a},
\skiplinehalf
\supit{a}University of Michigan, Ann Arbor MI, USA\\
}
\authorinfo{Further author information:\\
Send correspondence to M. Schubnell at schubnel@umich.edu\\}
\begin{document} 
  \maketitle 

%23456789.123456789.123456789.123456789.123456789.123456789.123456789.1234567890
\begin{abstract}
Flux dependent non-linearity (reciprocity failure) in HgCdTe NIR detectors
with $1.7\,\mu$m cut-off was investigated. A dedicated test station was designed
and built to measure reciprocity failure over the full dynamic range of near
infrared detectors. For flux levels between 1 and 100,000 photons/sec
a limiting sensitivity to reciprocity failure of 0.3\,\%/decade was achieved.
First measurements on several engineering grade $1.7\,\mu$m cut-off HgCdTe
detectors show a wide range of reciprocity failure, from less
than 0.5\,\%/decade to about 10\,\%/decade. For at least two of the tested
detectors, significant spatial variation in the effect was observed. No
indication for wavelength dependency was found. The origin of reciprocity failure is
currently not well understood. In this paper we present details of our
experimental set-up and show the results of measurements for several detectors.

%%This document shows the desired format and appearance of a manuscript 
%%prepared for the Proceedings of the SPIE.  It contains general 
%%formatting instructions and hints about how to use LaTeX.  
%%The LaTeX source file that produced this document, {\tt article.tex}
%% (Version 3.3), provides a template, used in conjunction with
%%  {\tt spie.cls} (Version 3.3).  
\end{abstract}

%>>>> Include a list of keywords after the abstract 


%%%%%%%%%%%%%%%%%%%%%%%%%%%%%%%%%%%%%%%%%%%%%
\section{INTRODUCTION}
\label{sec:intro}  % \label{} allows reference to this section

%23456789.123456789.123456789.123456789.123456789.123456789.123456789.1234567890
Near Infrared detector technology has made great strides over the past two
decades and large format arrays with excellent performance are now commercially
available at a cost that makes it possible to instrument imagers with
many 100s of millions of pixels.
Substrate-removed devices extend the wavelength sensitivity of near
infrared (NIR) detectors into the UV and highly
integrated read-out ASICS\cite{Loose} provide compact front-end electronics. The advances
in detector technology make NIR detectors well suited for space-based
wide-field imaging instruments, which are critical for pursuing some of the
big scientific questions of our time. One of the most far-reaching 
problems in physics today is the lack of understanding of the nature of dark
energy. The investigation of dark energy is most efficiently pursued with
experiments that employ a combination of different
observational techniques, such as type-Ia supernovae observations, gravitational weak lensing
surveys, galaxy and galaxy cluster surveys, and baryon acoustic oscillation measurements.
Most of these approaches rely on photometric calibrations over a wide range 
of intensities using standardized stars and internal reference sources, and
hence on a complete understanding of the linearity of the detectors.
As part of the SNAP R\&D effort\cite{Bebek2007} the Michigan near infrared
group has performed
a comprehensive study of precision photometry on $1.7\,\mu$m cut-off HgCdTe
detectors~\cite{Correlated_gain_paper, Schubnell_SPIE_2006, Spotomatic_PASP, Brown_PhD}.
Most recently, we have studied the count-rate dependent detector non-linearity
that was first reported from the Near Infra-Red Camera and Multi-Object
Spectrometer (NICMOS) on the Hubble Space
Telescope (HST)~\cite{NICMOSoverview, NICMOS0502, NICMOS0601, NICMOS0602}.
The NICMOS instrument, installed onboard the HST during the second service
mission in 1997, employs three 256$\times$256 NIR detectors. These $2.5\,\mu$m
cut-off HgCdTe devices were fabricated by Rockwell Science Center (now Teledyne
Imaging Sensors). The company also supplied the 1024$\times$1024 $1.7\,\mu$m
cut-off HgCdTe detector for the WFC3 instrument\cite{Baggett} which was recently
installed on the HST during the final service mission as well as the majority
of the detectors acquired by the SNAP Collaboration during the R\&D program.

The NICMOS team concluded that the NICMOS detectors exhibit a significant
count-rate dependent non-linearity that strongly varies with
wavelength\cite{NICMOS0502}. This flux dependent non-linearity,
referred to here as ``reciprocity failure,'' is
distinctly different from the well-known total count dependent non-linearity
observed in near infrared detectors that is caused by saturation as the well is filled.
Reciprocity failure can be characterized by a power law behavior
over most\footnote{In some detectors that we measured, notably detector H2-236, which shows
strong reciprocity failure, it appears that the detector response becomes linear at
very low and very high flux levels.} of the dynamic range of a detector: for a
given total integrated signal a pixel's response to a high flux
is larger than for a low flux.
% over a
%longer time and vice versa. 
The deviation from a linear system is expressed as
fractional deviation per decade of total signal response.
For the NICMOS array it was reported that this non-linearity is
correctable~\cite{NICMOS0603}, with corrections known to within 10\%. Efforts
were made to improve the uncertainty in the flux-dependent non-linearity
correction from 10\% to 5\%~\cite{NICMOS0704}.

Reciprocity failure impacts photometry as residual pixel-level uncertainties
directly propagate to the estimated uncertainty on the derived magnitude.
Detailed knowledge of the degree of reciprocity failure for a detector will
impact the calibration strategy and the calibration devices needed in space.
A profound understanding of the cause of this effect could impact the
manufacturing process and possibly reduce or even eliminate this non-linearity
component.

%
%23456789.123456789.123456789.123456789.123456789.123456789.123456789.1234567890

\section{EXPERIMENTAL SET-UP} 
\label{sec:setup}
 
In order to quantify reciprocity failure in NIR detectors, the Michigan NIR
group designed and built a dedicated test station. 
Based on the measurements reported by the NICMOS team it was determined that
a sensitivity to reciprocity failure of at least 1\%/decade over the full dynamic
range of a typical NIR detector (dark current level to full well) had to be achieved.
During the design phase, several methods for measuring detector reciprocity failure
were evaluated. 
The most intuitive approach is to simply vary a precisely monitored
incident flux onto a detector. However, this is a
challenging measurement at the 1\% level. Photodiode responses may be flux dependent
as well, and effects like temperature dependence and light source stability need to
be taken into account. An alternate method that was explored is based on the precise
inverse square law dependence of the photon intensity received at the
detector from a constant light source at varying distances. The drawback of this
method is that it requires an adjustable accordion like baffling system installed
inside a cold dewar. Controlling reflections
inside the dewar in the NIR is already non-trivial for a stationary baffling system. The
complexities of an adjustable design are numerous and difficult to overcome and
thus judged to not be a practical solution.
\\
%% tabular environment useful for creating an array of images  
\begin{figure}[hbt]
\begin{center}
\includegraphics[width=0.70\linewidth]{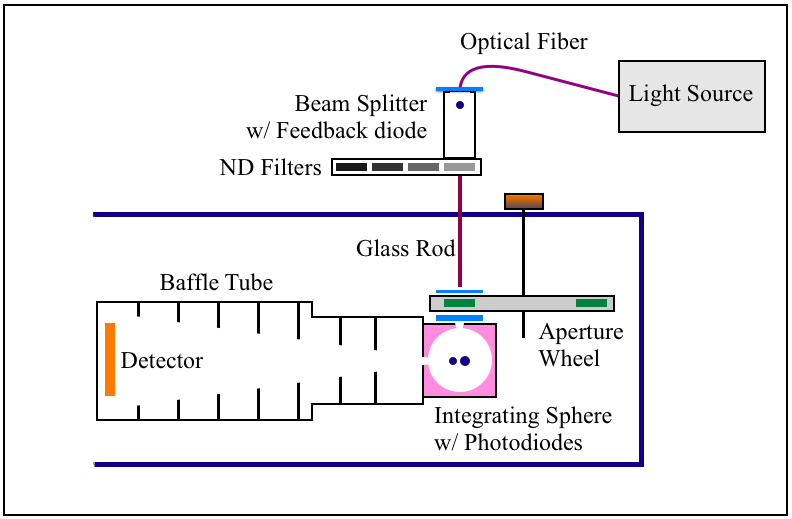}
\caption{\small Schematic overview of the set-up used to measure reciprocity
failure. The tested detector is well shielded from stray and background
light by a baffle system. A small integrating sphere is used to uniformly
illuminate the detector. A combination of neutral density filters and pinhole
apertures provides a large dynamic range in illumination intensity. Not shown
in the schematic is the liquid nitrogen vessel to which this extension is attached. 
}
\label{fig:setup}
\end{center}
\end{figure}

Ultimately a set-up was built that relies on a fixed geometry
while varying the detector illumination through a combination of
neutral density (ND) filters and pinhole apertures, as
schematically illustrated in Fig.~\ref{fig:setup}.
A regulated precision light source placed
outside the dewar is connected via a liquid light guide to a glass rod that
illuminates a pinhole mounted on the aperture wheel inside the dewar. To
avoid stray light entering the dewar, the glass rod is surrounded by a
bellows that attaches to the cold shield and the aperture wheel. The detector
is illuminated by an integrating sphere, placed immediately below the
aperture wheel, with a fixed aperture and baffling. This produces an
illumination profile at the 
detector that is independent of intensity. The baffling tube,
located between the integrating sphere and the detector, prevents stray light
and reflected light from reaching the detector and keeps the illuminating geometry fixed.
A set of six pinhole
apertures at the input of the integrating sphere
combined with
neutral density filters at the
entrance of the dewar extension
allow a dynamic range in intensity of approximately $10^6$ to be covered. 
For the measurement of reciprocity failure knowledge of the
exact area of the pinholes is not critical. 
% as long as the ratio of
%intensities produced by aperture pairs can be calibrated accurately.
Likewise, knowledge of the exact optical densities and the spectral dependence of the neutral density filters is also not
essential.

\subsection{Illumination}

The detector inside the dewar is illuminated by one of two controlled
light sources: a 50 W Quartz-Tungsten-Halogen (QTH) lamp or alternatively a 790\,nm diode laser.
Light from the QTH light source is guided by means of a liquid light guide (Newport 77634)
to a 70/30 beam splitter for feedback diode pick-up. The feedback diode is connected to the QTH lamp control electronics. This stabilizes the QTH light source to better than 0.1\%.
A filter stack in front of the beam splitter provides for pass band selection. Depending on the wavelength selected for the measurement either a 900\,nm long-pass filter or a combination of an 1100\,nm short-pass filter and an 1000\,nm short-pass filter (to improve out-of-band blocking) are inserted into the light path. The pass filter is then followed by one of four band-pass filters\footnote{The following band-pass filters  are used: 700 nm central wavelength, 80\,nm wide; 880\,nm, 50 nm wide; 950\,nm, 50\,nm wide and 1400\,nm, 80\,nm wide)}. Following the splitter, the re-focussed light beam passes through a filter slide, housing a selectable set of neutral density filters with optical densities 0,1,2, and 3. The connection from the warm optics into the dewar is made by a glass rod. Light from the glass rod is then incident on the selected aperture inside the aperture wheel. The aperture wheel has a total of 8 positions, six of which house pinholes ranging from 30\,$\mu$m to 6\,mm (30\,$\mu$m, 100\,$\mu$m, 330\,$\mu$m, 1\,mm, 3.3\,mm, 6\,mm), one position completely blocks the light and one position is fully open with no aperture.  

The pinhole illuminates the entrance port of 
an 2-inch integrating
sphere (SphereOptics SPH-2Z-4) as shown in Fig.~\ref{fig:setup}. An optional short-pass cold filter (Asahi YSZ1100) between two diffuser plates just in front of the integrating sphere is used for measurements below 1000\,nm. The inside of
the integrating sphere is coated with polytetrafluoroethylene (PTFE) based material
providing good reflectivity at NIR wavelengths.

\subsection{Monitoring Photodiodes}

The reciprocity set-up was optimized for measurement of candidate NIR detectors for JDEM\footnote{The Joint Dark Energy Mission, currently formulated as a joint strategic mission by NASA and the Department of Energy. SNAP, a dark energy mission concept has been incorporated into the design effort for JDEM.}. At the time of the first reciprocity measurements those detectors were substrate removed HgCdTe devices from TIS with a high wavelength cut-off of $1.7\,\mu$m. The process of substrate removal significantly improves the response between 400\,nm and 800\,nm. 
Therefore two monitor diodes, an InGaAs diode and a Si diode, were selected for good wavelength coverage. They are mounted to an open port of the integrating sphere and are read out in parallel. The NIR monitoring photodiode is a blue extended InGaAs PIN diode (Hamamatsu Photonics G108799-01K) with an effective area of 0.785 $mm^2$ and spectral response range of 0.5\,$\mu$m to 1.7\,$\mu$m. For improved sensitivity in the visible, a Si diode (Edmund Optics 53371) with an effective area of 5.1\,$mm^2$ and spectral response between 0.5\,$\mu$m and 1.1\,$\mu$m was used.
The two monitoring diodes are mounted adjacent to each other to an port of the integrating sphere as shown in Fig.\ref{fig:setup}.

Photodiode currents are monitored by two Keithley 6485 pico-ammeters
which are read out through a GPIB interface by the data acquisition
computer. For stable performance the pico-ammeter is turned on
at least 1 hour prior to every series of measurements. Typical photodiode currents are of order
10\,pA to 1\,$\mu$A.
%$^{-11}$\,A--10$^{-6}\,$A.
An accurate measurement requires multiple
samples. This is achieved by operating the pico-ammeter in sampling mode
and by averaging over 10 such samplings. 

Our measurement technique relies on the fact that the monitoring photodiodes are linear in their response or that any deviation from linearity is well known. Using a set of ND filters which were calibrated in a spectrometer-based test set-up for the measurement of astronomical filters, linearity tests were performed with both diodes. From those measurements we conclude that the diode response of each individual diode is linear to at least 1\% (at 270\,K) and that the relative linearity between the two diodes is better than 0.1\% over the full dynamic range used during the reciprocity measurements. Currently we can not verify diode linearity to better than 1\% but detailed linearity studies of Si and InGaAs photodiodes demonstrate linear behavior in those devices with uncertainties of about 0.2\% to 0.8\%\cite{Budde, Yoon}. For the large reciprocity failure observed in two of the tested devices (see Sec.\ref{sec:H2-236}) such a small potential systematic uncertainty does not alter the result of the measurement by much. 

%-------------
\begin{figure}[h]
\begin{center}
\begin{tabular}{c}
\includegraphics[width=0.50\linewidth]{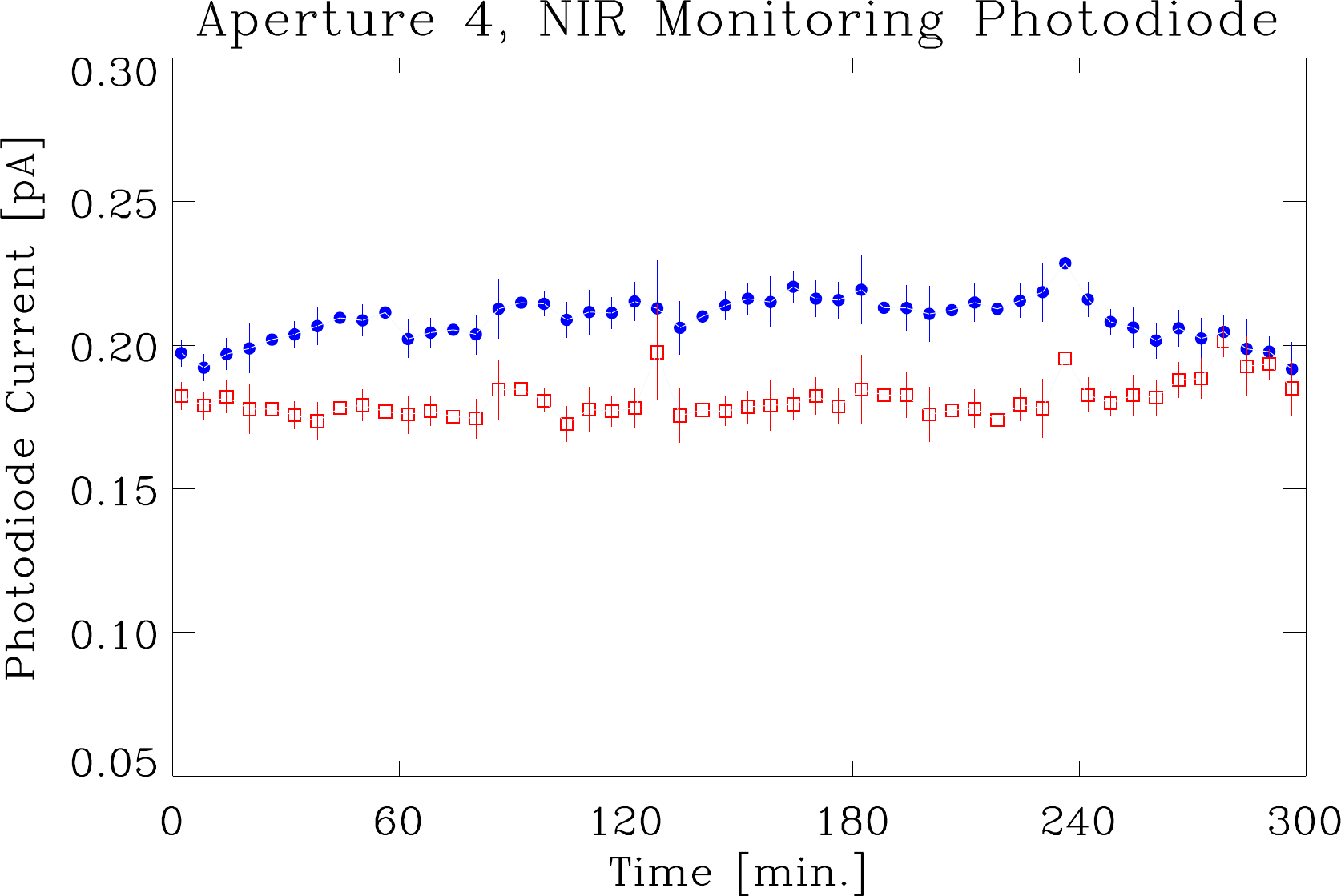}
\end{tabular}
\end{center}
\caption[photodiode]
{\label{fig:photodiode}
NIR monitoring photodiode current as a function of time. The blue circles show the time averaged dark corrected current registered during a reciprocity measurement extending over 5 hours. Such a long exposure is typical at the lowest illumination level (equivalent to a few electrons/second/pixel at the detector). The red squares show the same diode measurement corrected for fluctuations of the pico-ammeter. At the lowest setting the pico-ammeter drifts at the 10\% level. Those variations are tracked by a referencing photodiode connected to the pico-ammeter and can be removed from the data. Each data point in the figure represents the average of 500 photodiode current samples. The residual variation in the current measurement is dominated by statistical fluctuations and the variance of the mean improves linear with the number of measurements in the exposure.\\
}
\end{figure} 
%-------------

\subsection{Cryogenic System}

Prototype NIR devices are characterized at a baseline
temperature of 140\,K in an 8-inch dewar manufactured by IR Labs. The hold time of the system is typically 6--8\,hours, longer than the longest sampling sequence which takes about 5 hours to complete. This guarantees that measurements are not disrupted by the liquid nitrogen refill process.
For all measurements, the NIR detector is mounted to a fixed copper
heater plate which is weakly coupled to the liquid nitrogen
reservoir and thermally stabilized.
The cool-down and warm-up ramp of 1K/min as well as temperature stabilization of the NIR detectors
at the operating temperature is controlled and monitored by a precision
temperature controller (Lakeshore 330). 
With the
temperature of the detector held constant at $140\,$K, the illumination
system inside the dewar is allowed to cool down to about $150\,$K at the integrating sphere over a time
period of about eight hours. This is much colder than required to suppress thermal background radiation in the  $1.7\,\mu$m detector material. A second temperature control loop is used to eliminate temperature dependence in the
response of %of 0.2\% per K, as shown in the right panel of Fig.~\ref{fig:cooldown},
the monitoring photodiodes. The two photodiodes are always temperature stabilized to a common temperature of $270\,$K.

\subsection{Read-out and Control Electronics}

For detector read-out and control, a commercially available data acquisition system
from
Astronomical Research Cameras (ARC) was used. In this system,
32 channels of parallel read-out are available from four 8-channel
infrared video processor boards combined with infrared clock driver boards and
250\,MHz timing and PCI cards. This read-out electronics is described in
detail in Leach \& Low~\cite{Leach}.
Data are stored in FITS format for subsequent analysis.
In the current set-up no shutter was employed and thus each detector pixel starts to integrate signal right after reset. Consequently, the shortest ``illumination time'' is determined by the amount of time it takes to read the array. In the default clocking mode (100\,MHz) the read-out of the whole array takes 1.418 seconds. To reduce the illumination time for this test, only a partial strip of the detector, 300$\times$2048 pixels in size, was read. This reduced the read-out time to 211 milliseconds.

\section{System Performance and Results} 

Several 1.7\,$\mu$m HgCdTe focal plane arrays (FPAs) from Teledyne Imaging Sensors (TIS), procured by the SNAP
program, were characterized with our reciprocity set-up.
%Here we present an overview of the results from these measurements.

The FPAs are hybridized detectors consisting of a highly integrated CMOS multiplexer
and a layer of infrared sensitive detector material.
Photon conversion takes place in a very thin layer of  HgCdTe, about 5 -- 10 $\mu$m thick
(typically grown on a much thicker substrate layer) with metallized
contact pads defining the active area. The accumulation of
photogenerated electron-hole pairs on the isolated photodiode causes a
decrease in reverse bias which is sensed by a MOSFET source
follower. The multiplexer is an array of discrete read-out transistors
and, unlike a conventional CCD, can be read non-destructively.
Detector layer and multiplexer are indium bump-bonded and mounted to a pedestal which
equilibrates the temperature across the FPA.

The SNAP project pursued a strong detector procurement and
development program for 1.7\,$\mu$m HgCdTe FPAs with the goal of producing
a low read-noise, high quantum efficiency (QE) detector suitable for the proposed SNAP instrument\cite{schubnell2004}. Several production lots were produced by TIS, with fabrication based on the Hubble Space Telescope's Wide
Field Camera 3 (WFC3) development of 1k$\times$1k HgCdTe. Low read-noise, high QE and substrate removal were addressed during different material growth runs. The detectors for which reciprocity failure is reported here came from lot \#4 (H2-102), lot \#5 (H2-142) and lot \#6 (H2-236 and H2-238).
%%%% CHECK WITH CHRIS THAT THIS IS RIGHT %%%%%%%%%%%%%%%%%
 
\subsection{Test and Analysis Procedure}

\begin{figure}[h]
\begin{center}
\begin{minipage}[b]{0.45\linewidth}
\begin{center}
\begin{tabular}{c}
\includegraphics[width=\linewidth]{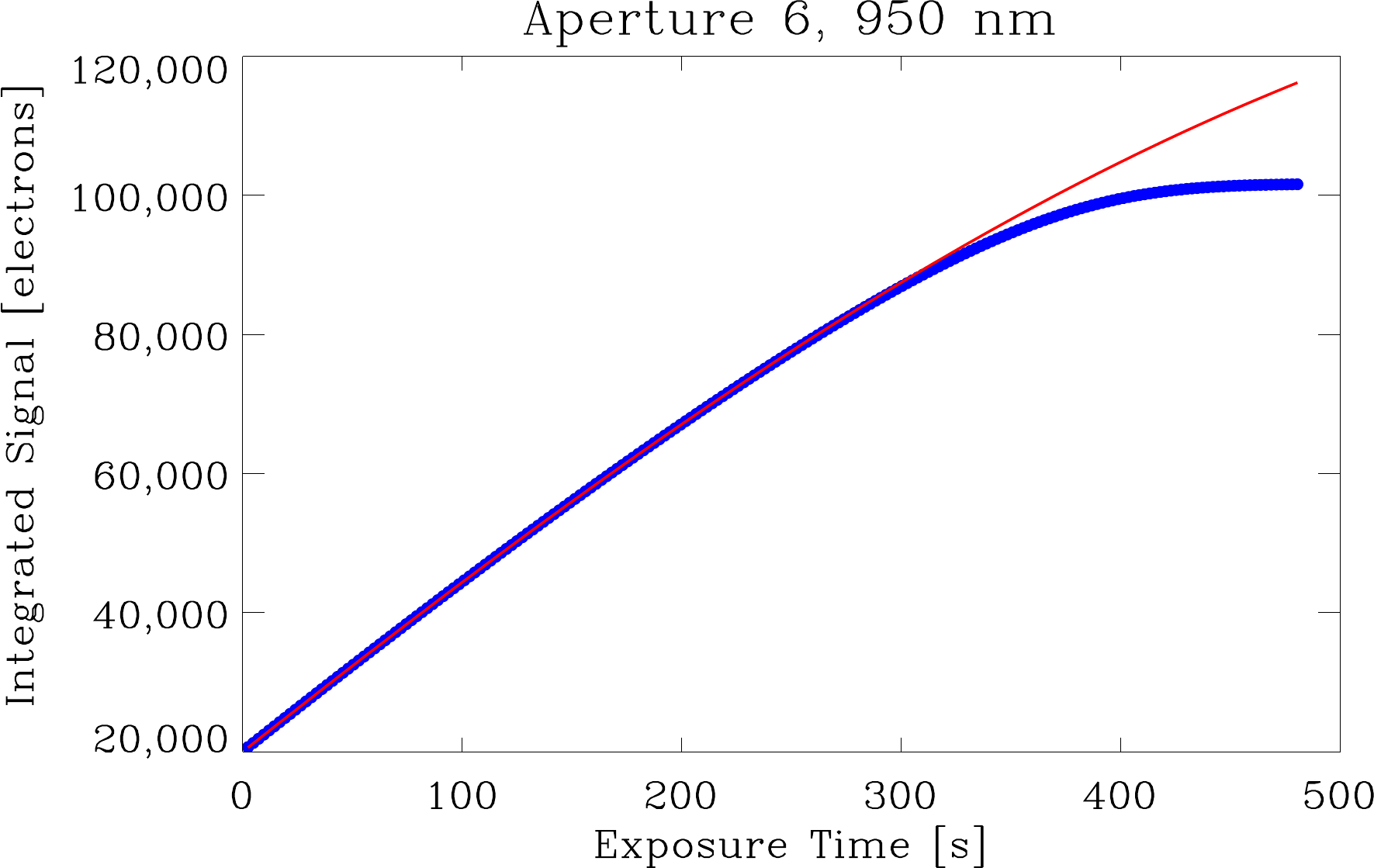}
\end{tabular}
\end{center}
\end{minipage}
\hspace{0.4cm}
\begin{minipage}[b]{0.45\linewidth}
\begin{center}
\begin{tabular}{c}
\includegraphics[width=\linewidth]{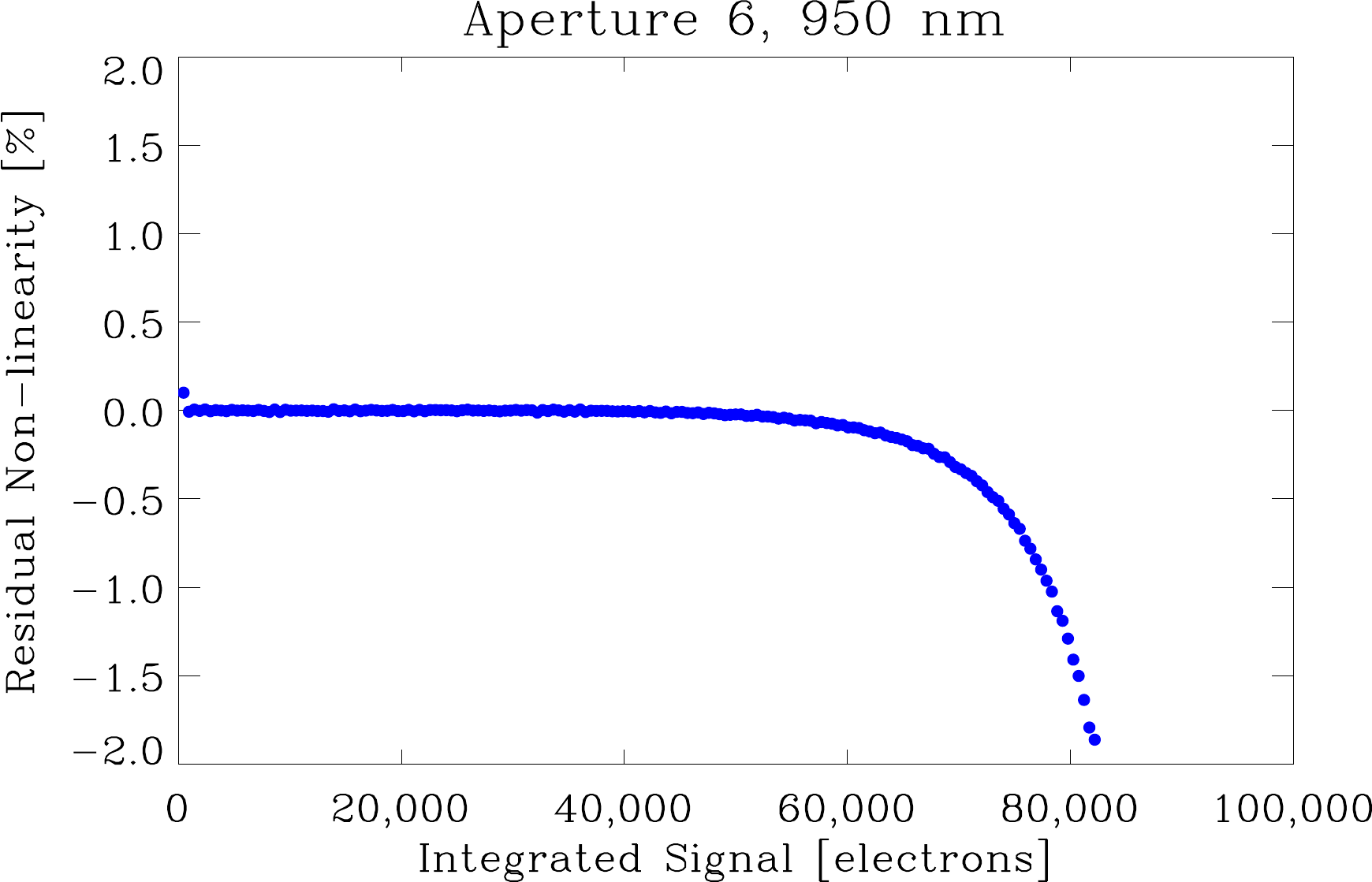}
\end{tabular}
\end{center}
\end{minipage}
\caption[total_signal]{The integrated signal in a HgCdTe detector as a function of time (left panel). Signal non-linearity of the detector as the well is filled is well understood and can be parameterized. After applying the correction, signal non-linearity is reduced to 0.1\% for signal levels below 60\% of the well depth (right panel). Up to 80 \% of the well depth the non-linearity is still below 1\%. }
\label{fig:total_signal}
\end{center}
\end{figure}

During a typical reciprocity measurement the detector is read out in ``up the ramp'' sampling mode (SUR) with up to 200 frames read during an exposure. Baseline dark ``bias frames'' are obtained with the illumination blocked from the detector at minimal exposure time. These frames establish the baseline well depth and serve as reference for the SUR exposure. 
In addition ``matched darks'' are obtained for every SUR sequence. Measurement conditions for the matched darks are in every way identical to the reciprocity measurement conditions but are taken with the light source turned off. 
The signal non-linearity (i.e., the non-linearity of detected counts with 
total incident flux) is characterized by measuring the detector response as a function of the integration time at a fixed incident flux. The integrated signal, $S$, in the detector is parameterized as $S(t,F)=\int{c(t,F)\times\epsilon(S)}$, where $c(t,F)$ represents the detector count-rate as a function of time $t$ and incident flux $F$, and $\epsilon(S)$ takes into account the detection efficiency.
As can be seen in Fig.\,\ref{fig:total_signal}, applying this parameterization describes the observed behavior well. After the correction, the signal non-linearity for signals below 60\% of the well depth is less than 0.1\%. Matched darks and signal non-linearity data are then included in the fit procedure used in calculating the NIR detector response.

During the reciprocity measurement the photodiode currents are recorded for each frame in the sample. From this the ratio of detector signal and diode current is computed and combined with the matched darks and the signal non-linearity data. A fit procedure calculates the normalized flux ratios shown in Figs.\,\ref{fig:rec_failure_102}, \ref{fig:rec_failure_142} and \ref{fig:rec_failure_236}. Normalized flux ratios are obtained at different illumination intensities and at different wavelengths.
At wavelengths below 1000\,nm current readings from the Si diode and above 1000\,nm data from the InGaAs diode are used for calculating the flux ratios.

%{\it when is the well-fill non-linearity taken out?}

\subsection{Results and Discussion}

Reciprocity measurements were performed on four different HgCdTe detectors. For all measurements presented here the NIR detector temperature was held constant at 140\,K. Measurements were made with the QTH illumination system at wavelengths of 700\,nm, 880\,nm, 950\,nm and 1400\,nm and with the laser at 790\,nm.

\subsubsection{H2-102}

Detector H2-102 is a well performing engineering grade device from the SNAP
R\&D program.  The detector is substrate
removed and has an anti-reflective coating. QE is over 90\% from 0.9\,$\mu$m -- 1.7\,$\mu$m and the response extends into the 
visible; the measured QE at 0.45\,$\mu$m is about 40\%. Dark current and read-noise performance is good, with a measured Fowler-16 noise of $10\,e^-$ for a $300\,$s exposure at $140\,$K.
Unlike all the other 
devices for which detector reciprocity was measured, this array was mounted onto a Molybdenum pedestal. 

\begin{figure}[h]
\begin{center}
\begin{minipage}[b]{0.45\linewidth}
\begin{center}
\begin{tabular}{c}
\includegraphics[width=\linewidth]{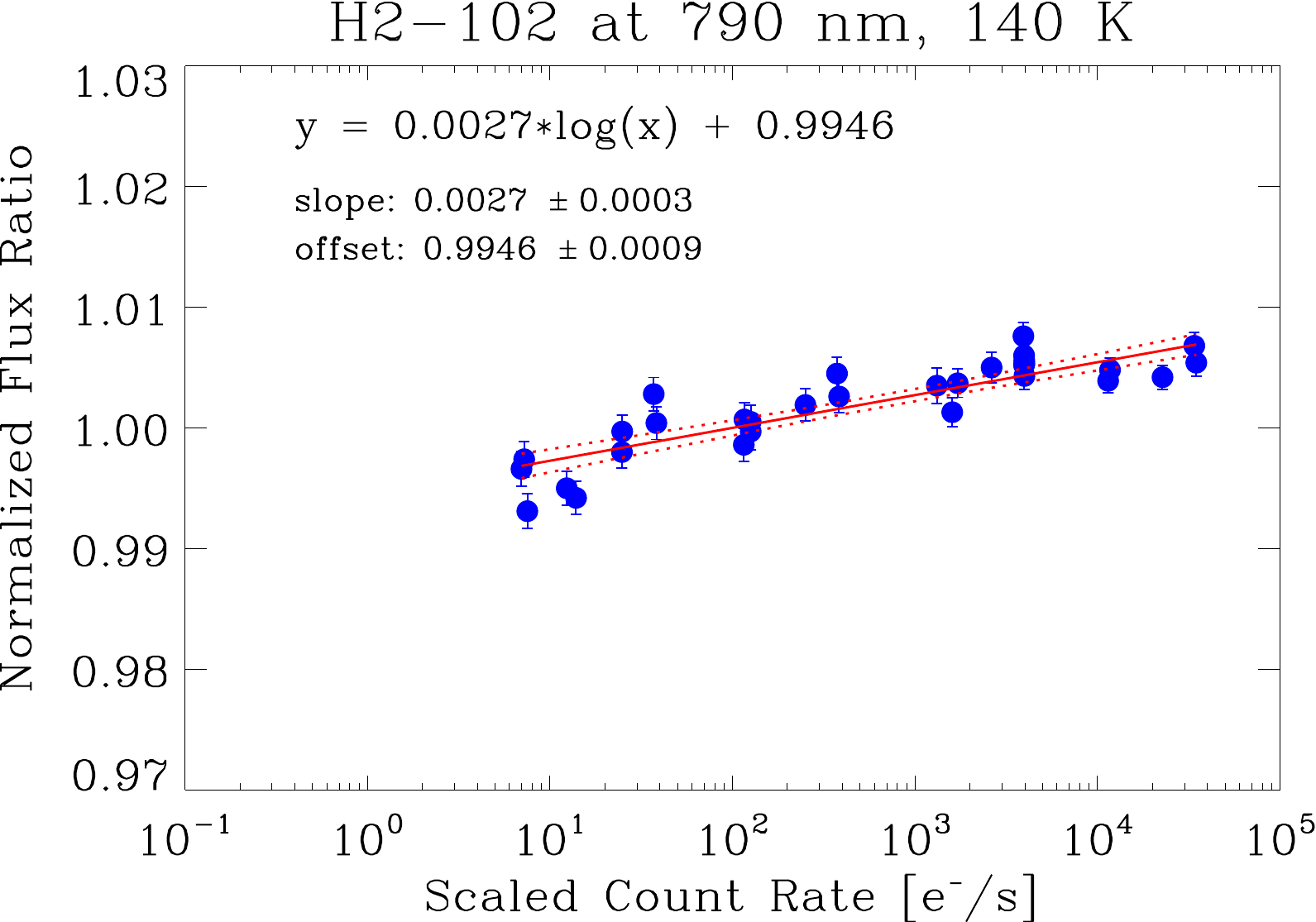}
\end{tabular}
\end{center}
\end{minipage}
\hspace{0.4cm}
\begin{minipage}[b]{0.45\linewidth}
\begin{center}
\begin{tabular}{c}
\includegraphics[width=\linewidth]{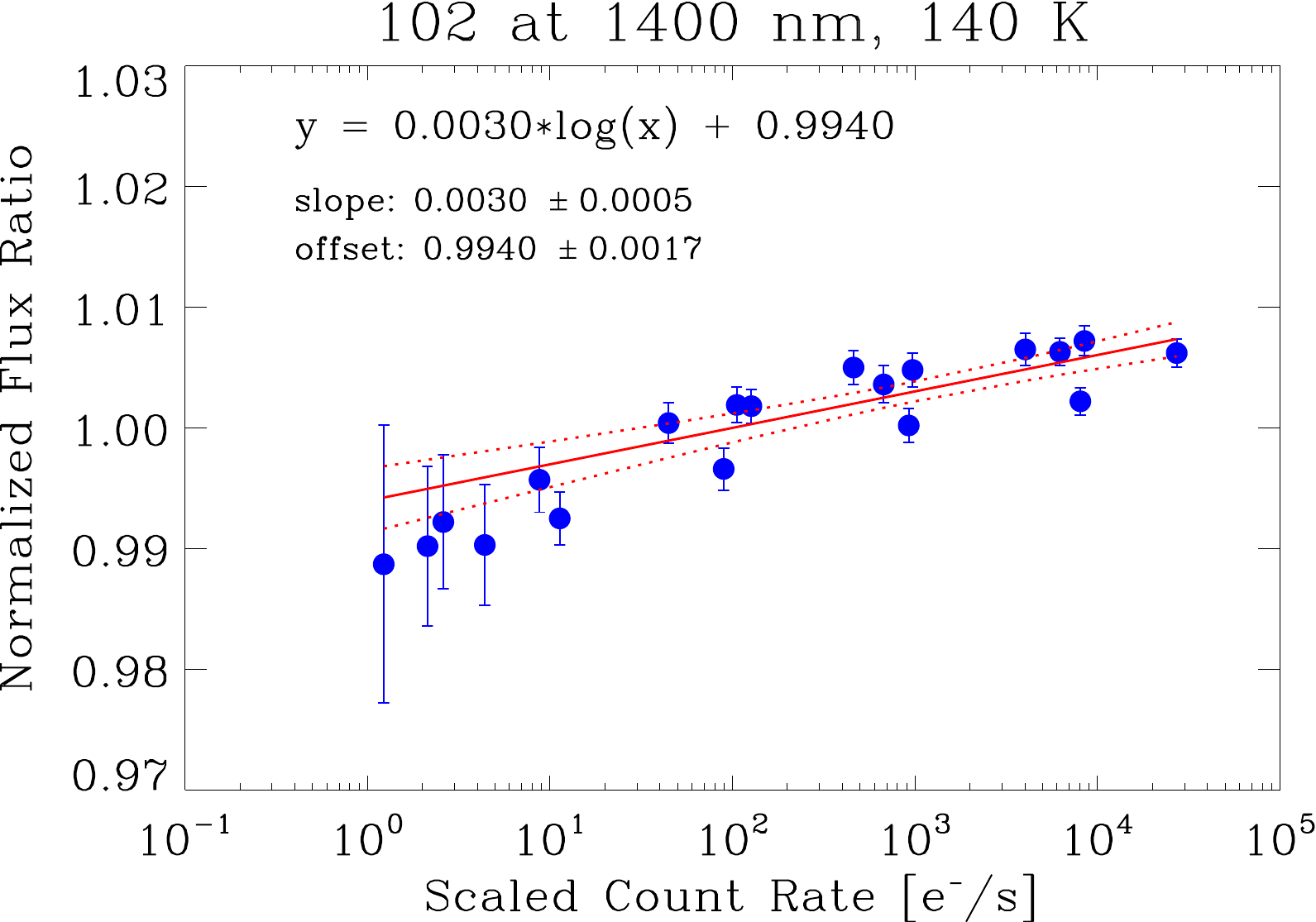}
\end{tabular}
\end{center}
\end{minipage}
\caption[rec_failure_236]{Reciprocity failure measured in device H2-102 at two different wavelengths (790\,nm and 1400\,nm). The NIR detector count rate is scaled relative to the the photodiode current to remove flux dependency from the horizontal axis (``Scaled Count Rate''). In each panel a solid line indicates the best power law fit to the data points. The dotted lines represent the 1$\sigma$ bounds.}
\label{fig:rec_failure_102}
\end{center}
\end{figure}

H2-102 shows low reciprocity failure (see Fig.\,\ref{fig:rec_failure_102}). 
A power law fit (linear in log illumination) describes the effect well.
Measurements were performed at 5 different wavelengths (700\,nm, 790\,nm, 880\,nm, 950\,nm and 1400\,nm)
with no significant wavelength dependency observed. Measured values for the
reciprocity failure at the five wavelengths 
are (0.26$\,\pm\,$0.04)\,\%/decade, (0.27$\,\pm\,$0.03)\,\%/decade, (0.28$\,\pm\,$0.06)\,\%/decade, (0.21$\,\pm\,$0.04)\,\%/decade, and (0.30$\,\pm\,$0.05)\,\%/decade.
This contrasts the strong wavelength dependency for 
reciprocity failure in all three NICMOS detectors.\\

%{\it How are the error bars calculated?}

\subsubsection{H2-142}
\begin{figure}[h]
\begin{center}
\begin{minipage}[b]{0.45\linewidth}
\begin{center}
\begin{tabular}{c}
\includegraphics[width=\linewidth]{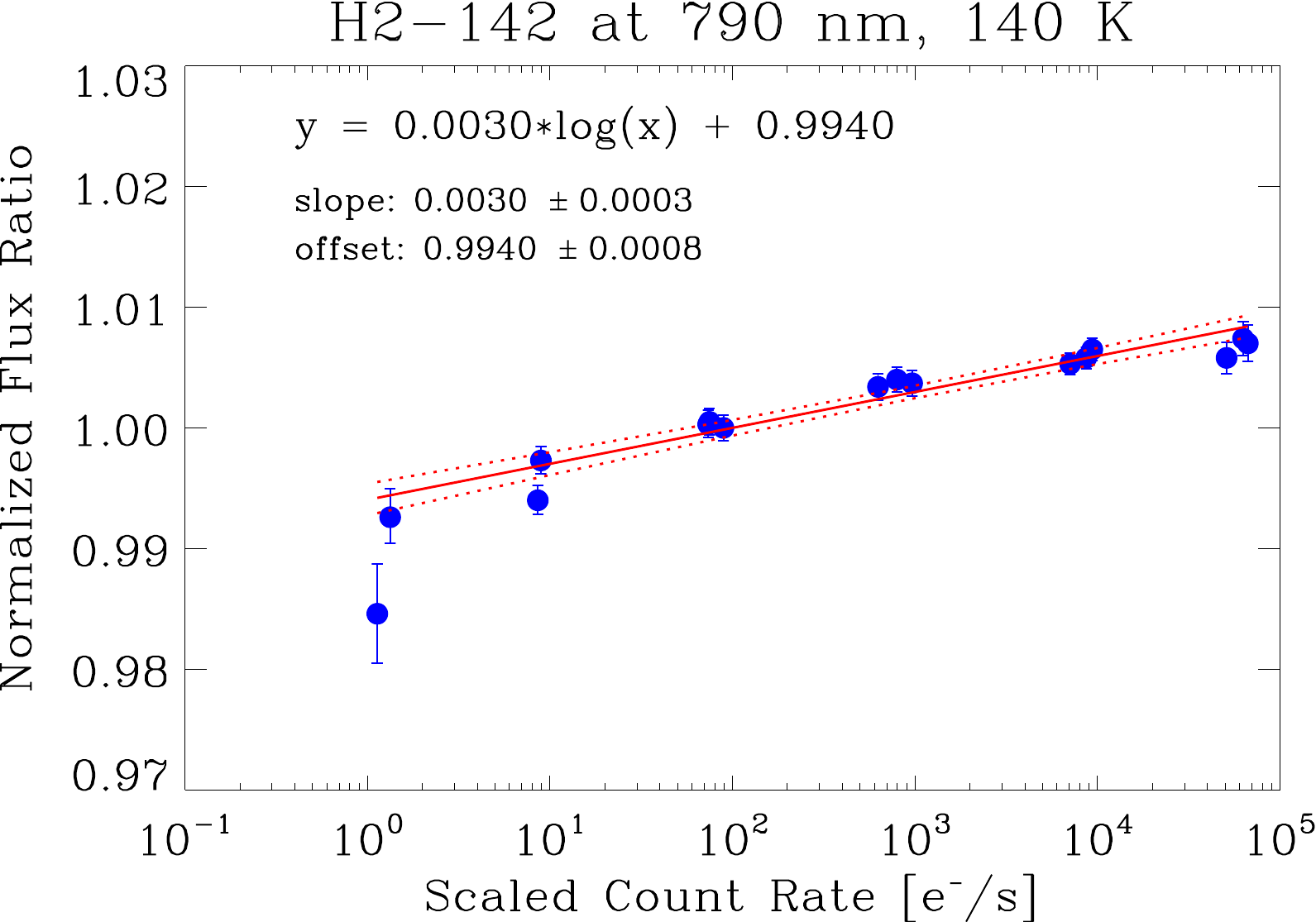}
\end{tabular}
\end{center}
\caption[rec_failure_236]{Reciprocity failure in device H2-142 at 790 nm.}
\label{fig:rec_failure_142}
\end{minipage}
\hspace{0.5cm}
\begin{minipage}[b]{0.35\linewidth}
\begin{center}
\begin{tabular}{c}
\includegraphics[height=5.5cm]{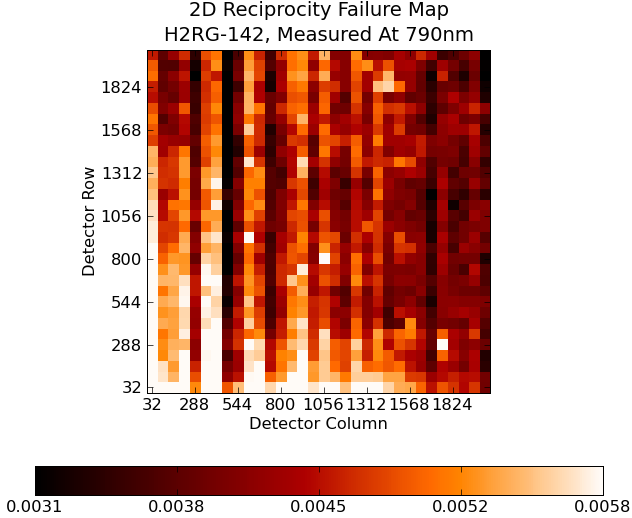}
\end{tabular}
\end{center}
\caption[rec_failure_142_map]{Spatial variation of reciprocity failure in H2-142.}
\label{fig:rec_failure_142_map}
\end{minipage}
\end{center}
\end{figure}

H2-142 is a device from the 5$^th$ manufacturing run
%%%%% CHECK THIS
for SNAP by TIS and was one of the first devices mounted onto a SiC pedestal, specifically developed for SNAP. 
%{\it(reference)}. 
The device performed well in our standard tests. CDS read noise is low (24.7\,$e^-$) and QE is approaching 90\,\% for the NIR region between 900 nm and 1700 nm. The measured reciprocity failure of (0.30$\,\pm\,$0.03)\,\%/decade (Fig.\ref{fig:rec_failure_142}) at 790 nm is remarkably consistent with the reciprocity failure measured for detector H2-102. During the standard ``strip'' measurement on one of the tested devices (H2-236), spacial structure in the reciprocity failure across the strip became apparent. This led us to perform a measurement with the full detector. Results from this measurement are shown in Fig.\,\ref{fig:rec_failure_142_map}. Each pixel in this map corresponds to a 64 pixel\,$\times$\,64 pixel area on the device. Reciprocity failure in H2-142 varies by about a factor of two, with the effect being strongest in one corner of the detector (bottom left in Fig.\,\ref{fig:rec_failure_142_map}). At lower illumination levels the detector appears more uniform.\\

%-------------
%-------------

\subsubsection{H2-236 and H-238}
\label{sec:H2-236}
The two detectors H2-236 and H2-238 are devices produced during the 6$^th$ manufacturing run for SNAP. In this run TIS addressed the issue of capacitive coupling by hybridizing the detector to a slightly modified version of their standard Hawaii-2 read-out multiplexer. Like H2-142 this detector is mounted onto a SiC pedestal. Manufacturing and processing of the detector followed largely the recipe that produced the excellent flight detector produced for the WFC3 team\cite{Baggett}. Dark current and read noise performance for both detectors is good, with a CDS read noise of 25.7\,$e^-$/s for H2-236 and 22.3\,$e^-$/s for H2-238. QE is acceptable but about 10\% lower than what had been achieved for previous detectors. While a slight improvement of the interpixel capacitive coupling was achieved, the most notable performance difference to previously produced devices was an increase in persistence by almost a factor of 10. 

\begin{figure}[t]
\begin{center}
\begin{minipage}[b]{0.42\linewidth}
\begin{center}
\begin{tabular}{c}
\includegraphics[width=\linewidth]{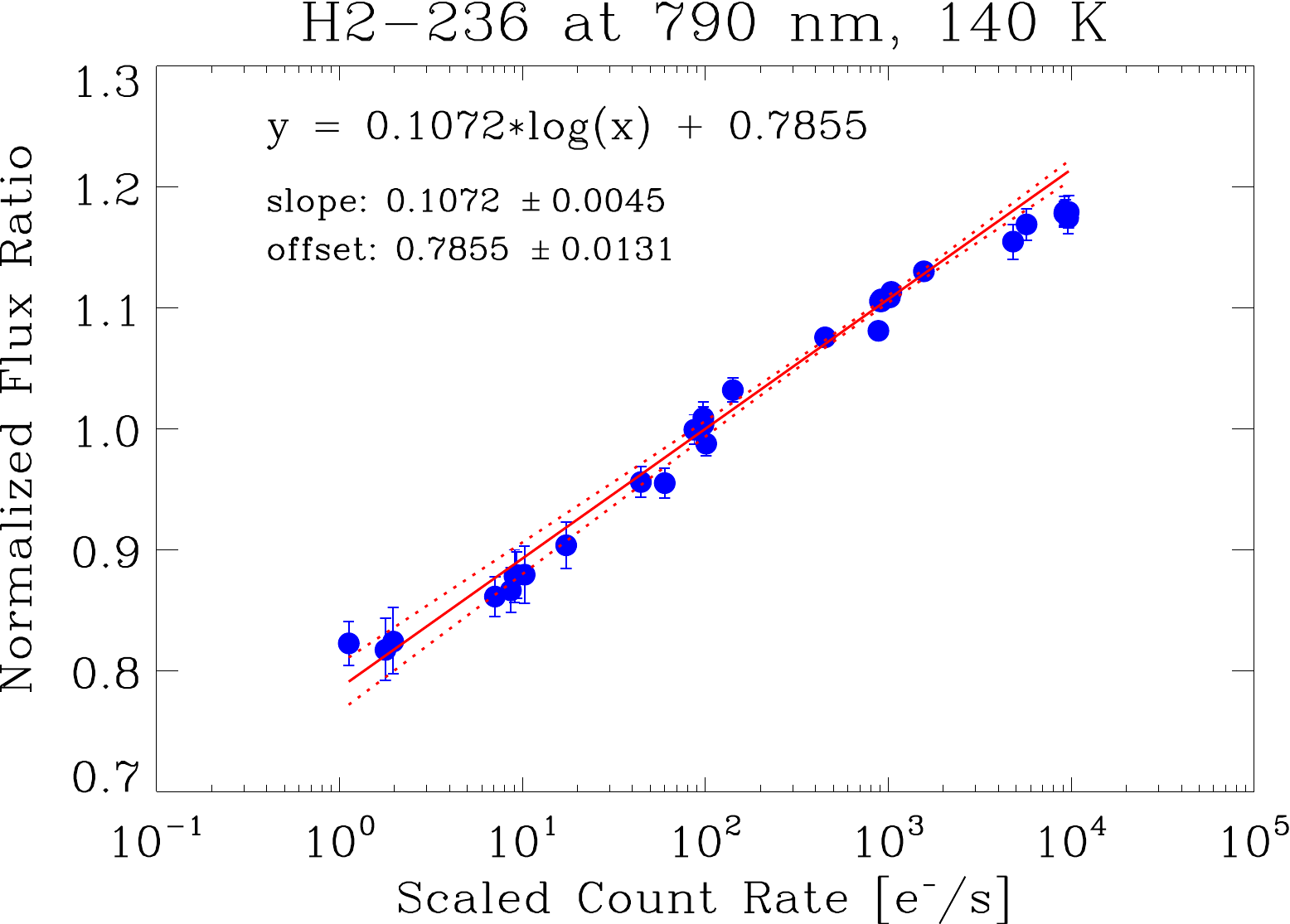}
\end{tabular}
\end{center}
\end{minipage}
\hspace{0.3cm}
\begin{minipage}[b]{0.42\linewidth}
\begin{center}
\begin{tabular}{c}
\includegraphics[width=\linewidth]{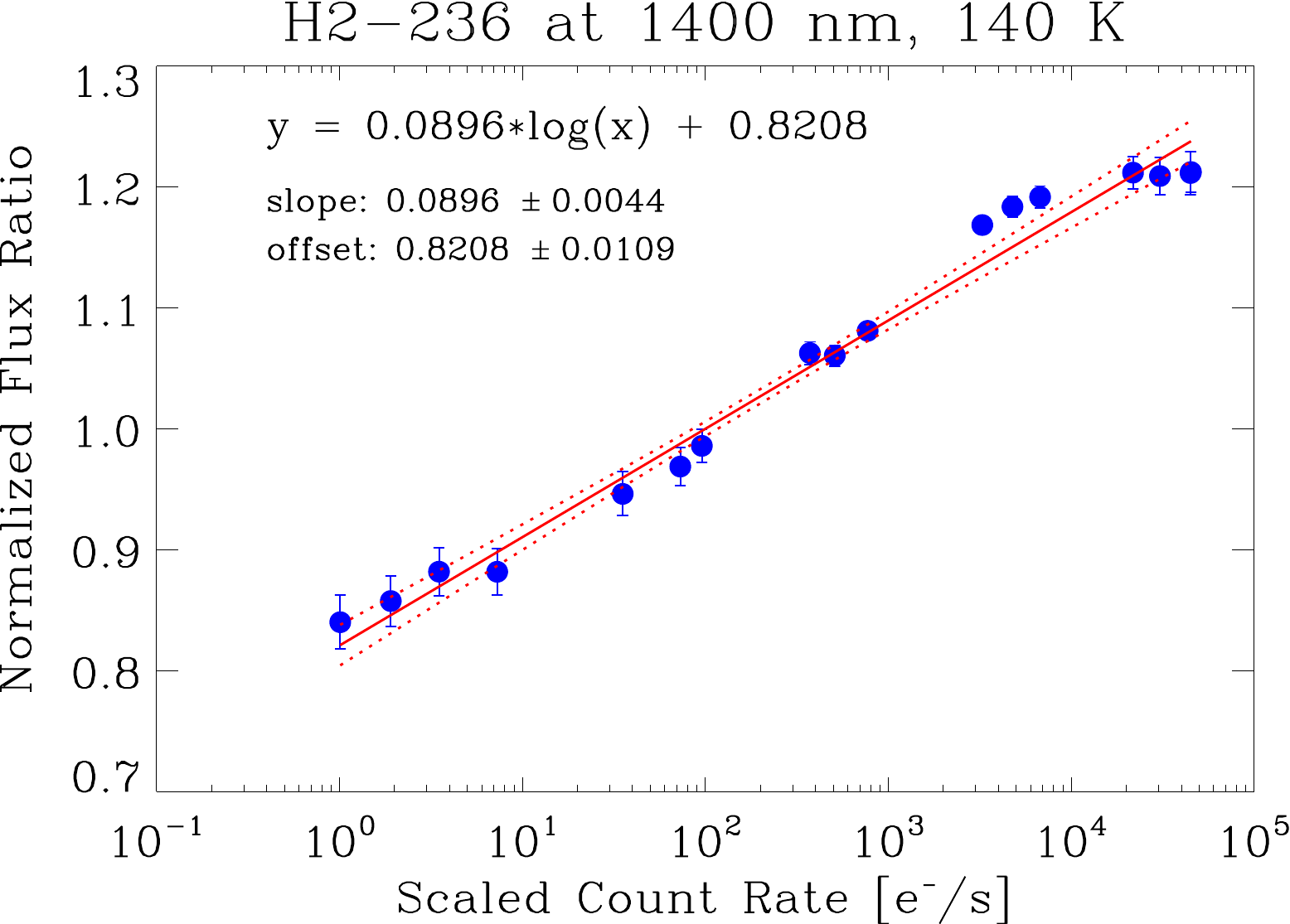}
\end{tabular}
\end{center}
\end{minipage}
\caption[rec_failure_236]{Reciprocity failure measured in device H2-236 at 790\,nm (left panel) and 1400\,nm (right panel). Note the dramatic change of scale in the vertical axis compared to Figs.\ref{fig:rec_failure_102} and \ref{fig:rec_failure_142}.}
\label{fig:rec_failure_236}
%\end{center}
\vspace{0.6cm}
%\end{figure}
%\begin{figure}[h]
%\begin{center}
\begin{minipage}[b]{0.42\linewidth}
\begin{center}
\begin{tabular}{c}
\includegraphics[width=\linewidth]{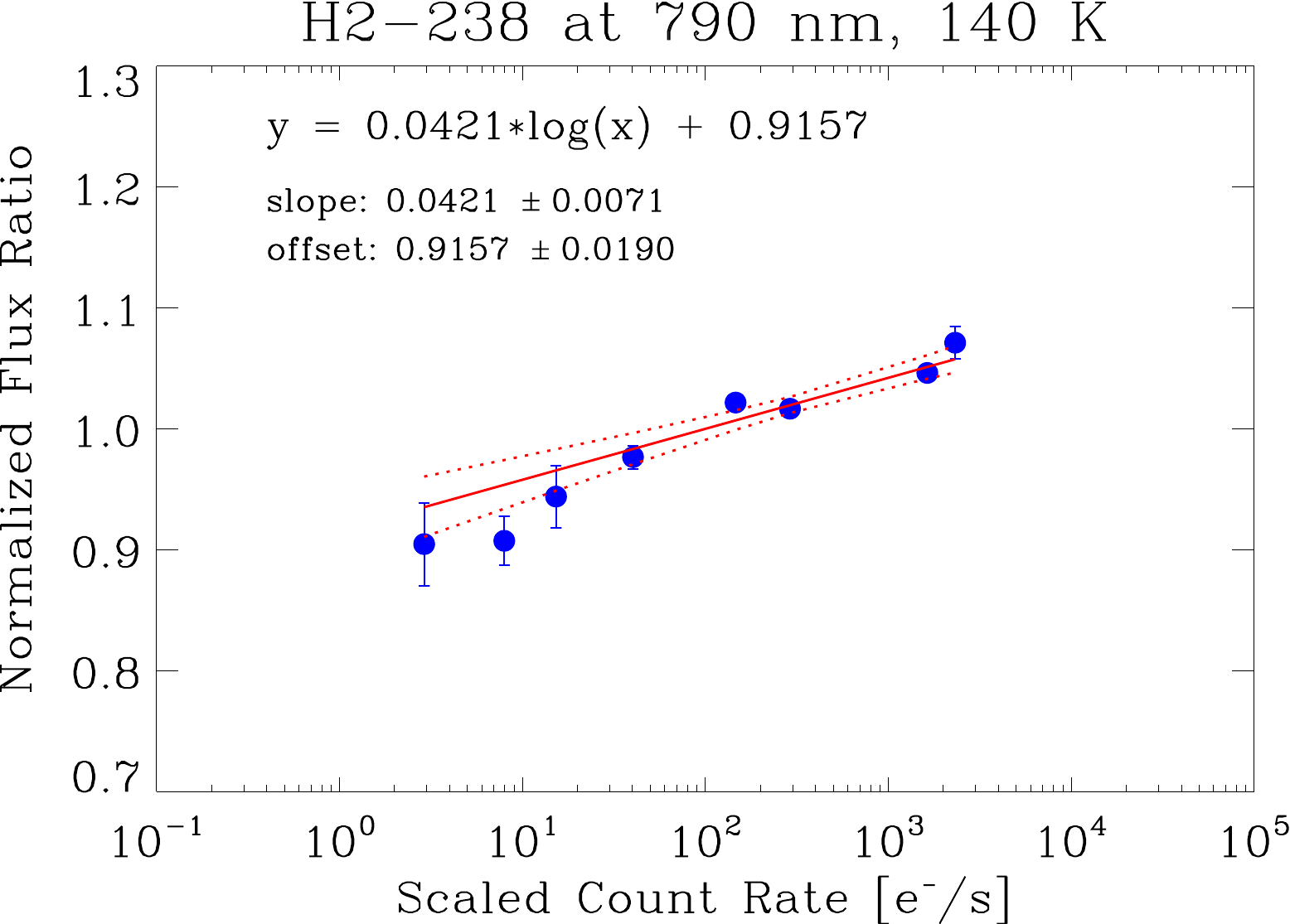}
\end{tabular}
\end{center}
\end{minipage}
\hspace{0.3cm}
\begin{minipage}[b]{0.42\linewidth}
\begin{center}
\begin{tabular}{c}
\includegraphics[width=\linewidth]{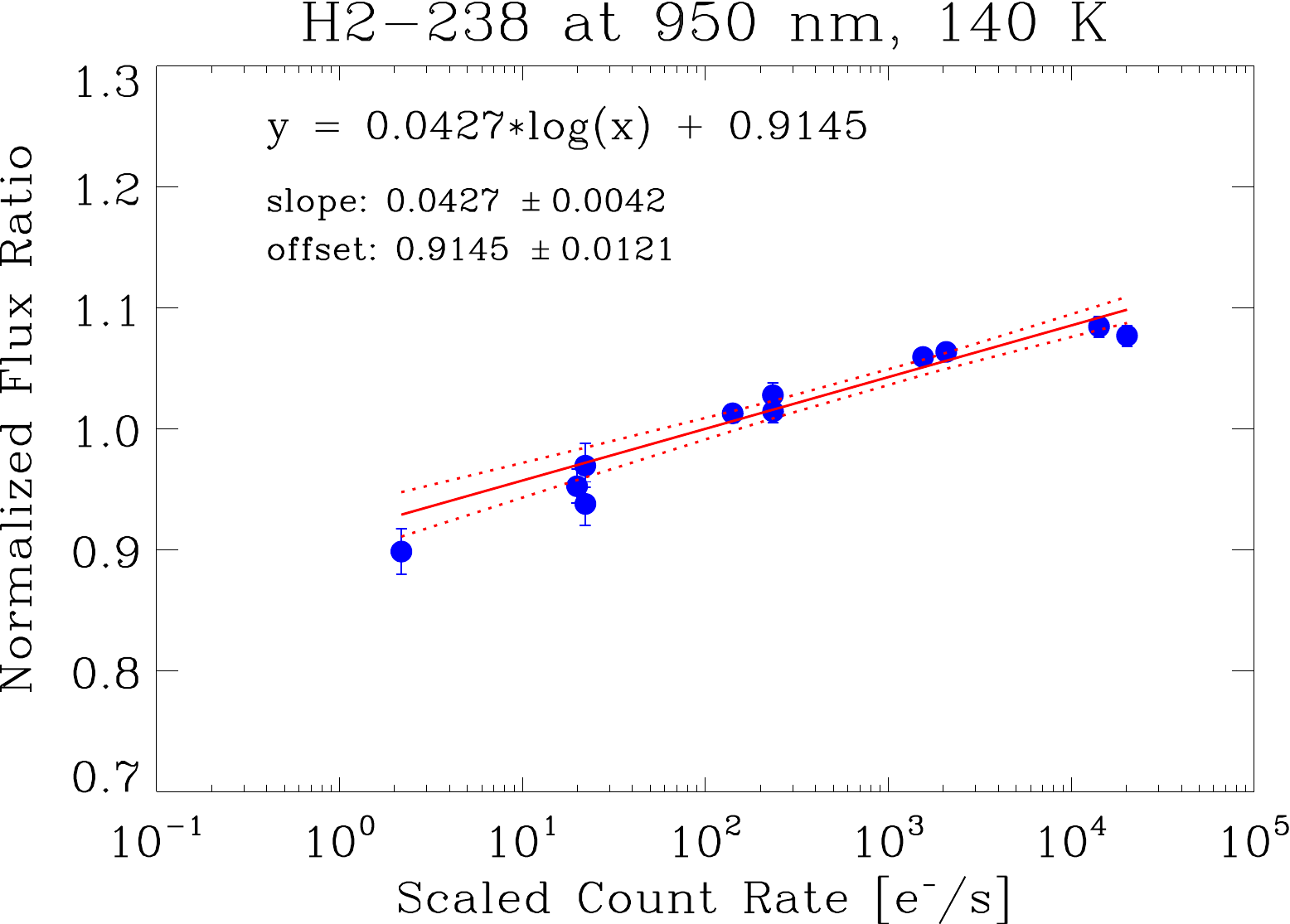}
\end{tabular}
\end{center}
\end{minipage}
\caption[rec_failure_238]{Reciprocity failure in device H2-238 at 790\,nm (full detector) and 950\,nm (strip).  }
\label{fig:rec_failure_238}
%\end{center}
%\end{figure}
\vspace{0.6cm}
%\begin{figure}[h]
%\begin{center}
\begin{minipage}[b]{0.42\linewidth}
\begin{center}
\begin{tabular}{c}
\includegraphics[height=5.5cm]{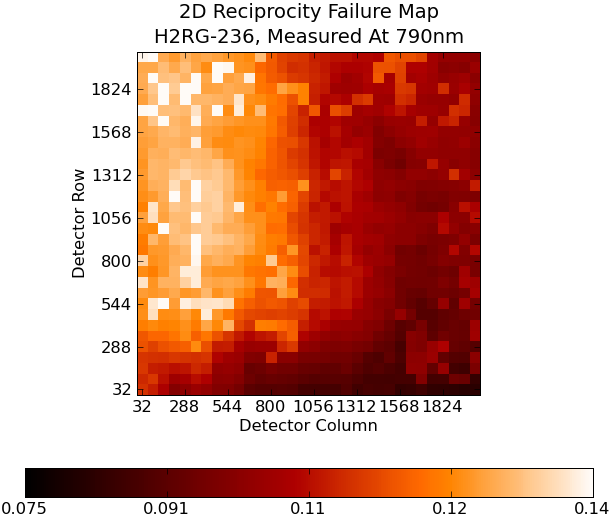}
\end{tabular}
\end{center}
\end{minipage}
\hspace{0.3cm}
\begin{minipage}[b]{0.42\linewidth}
\begin{center}
\begin{tabular}{c}
\includegraphics[height=5.5cm]{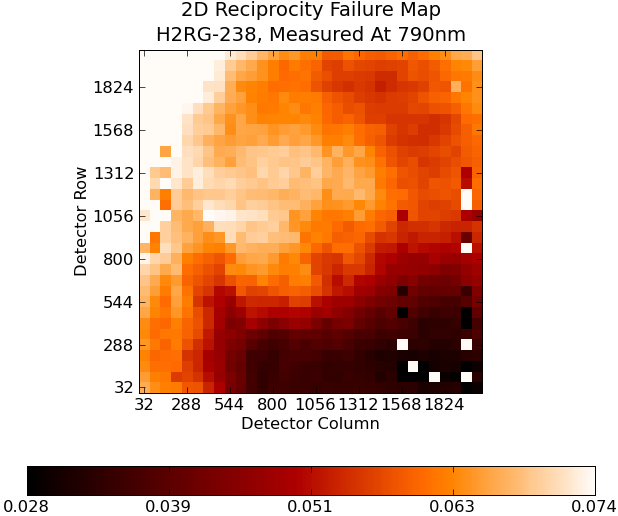}
\end{tabular}
\end{center}
\end{minipage}
\caption[rec_failure_236]{Strong spatial variation of reciprocity failure was observed in H2-236 and H2-238.}
\label{fig:rec_failure_spatial}
%\begin{center}
%\begin{tabular}{c}
%\includegraphics[width=0.31\linewidth]{rec_failure_map_236.png}
%\end{tabular}
%\caption[rec_failure_236]{Reciprocity failure measured in device H2-236.}
%\label{fig:rec_failure_236}
\end{center}
\end{figure}

%Combined with the very large reciprocity failure observed for these two devices, this indicates a common underlying cause, possible an increase in trap density. 
Again, a power law fit describes the effect well, although it appears that reciprocity failure ceases/saturates at low and high illumination levels, i.e. the detector behaves more linear in those regimes. For detector H2-236 a reciprocity failure of (10.79$\,\pm\,$0.36)\,\%/decade at 790\,nm for the full detector was measured. 
The error bars in this case are dominated by the detector gradient in reciprocity failure.
At 950\,nm a reciprocity failure of (11.16$\,\pm\,$0.32)\,\%/decade was measured. This is the most severe effect observed in any of the measured devices so far. Additionally, significant spacial structure is observed in both detectors (Fig.\,\ref{fig:rec_failure_spatial}).%\begin{figure}[hbt]
%\begin{center}
%\begin{tabular}{c}
%\includegraphics[width=0.99\linewidth]{rec_fail_236_diff_wavelength.png}
%\end{tabular}
%\caption[rec_failure_236]{Reciprocity failure measured in device H2-236.}
%\label{fig:rec_failure_236}
%\end{center}
%\end{figure}
\newpage

\subsection{Summary}

\begin{figure}[h]
\begin{center}
\begin{minipage}[b]{0.44\linewidth}
\begin{center}
\begin{tabular}{c}
\includegraphics[width=\linewidth]{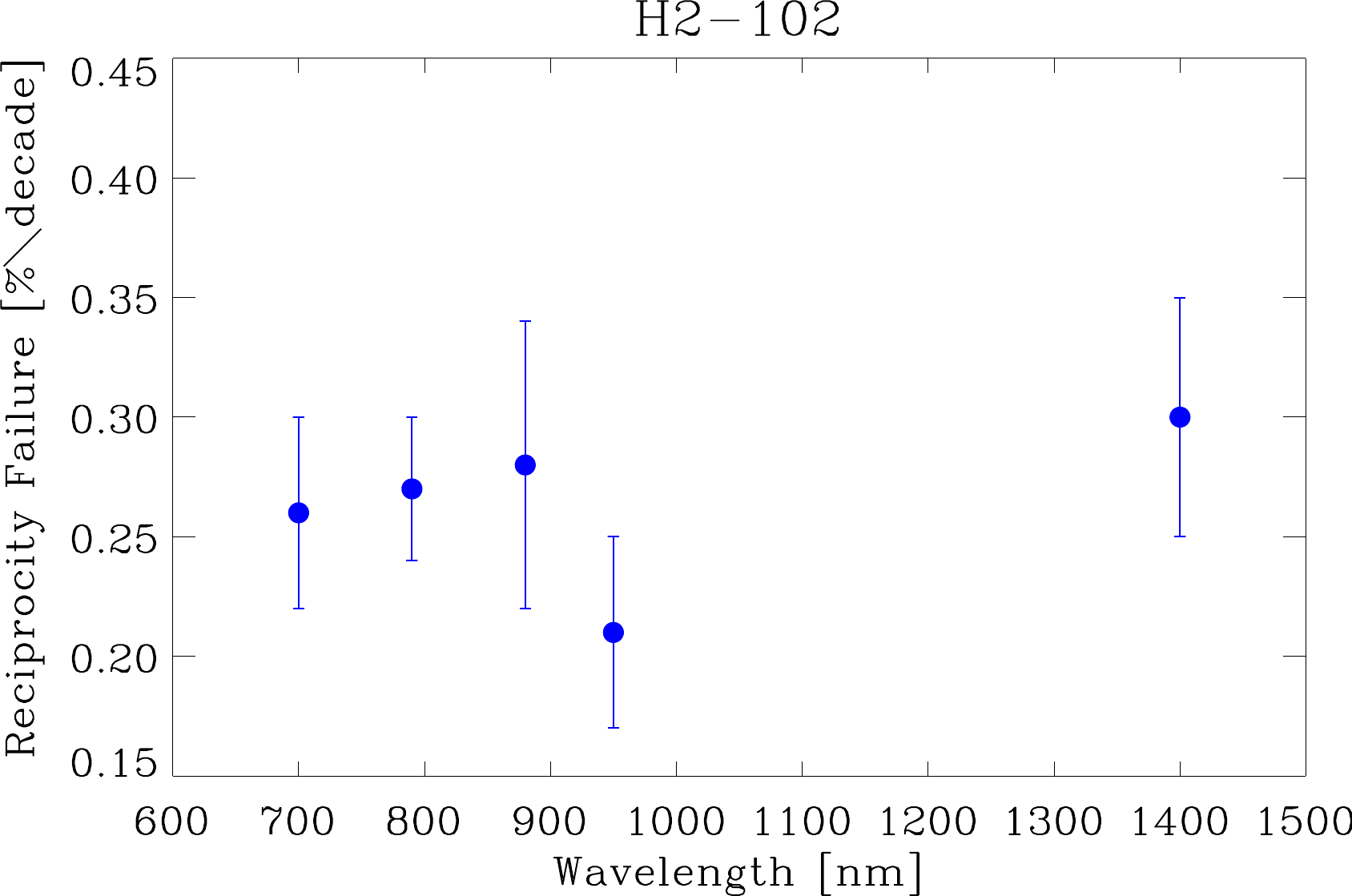}
\end{tabular}
\end{center}
\end{minipage}
\hspace{0.5cm}
\begin{minipage}[b]{0.44\linewidth}
\begin{center}
\begin{tabular}{c}
\includegraphics[width=\linewidth]{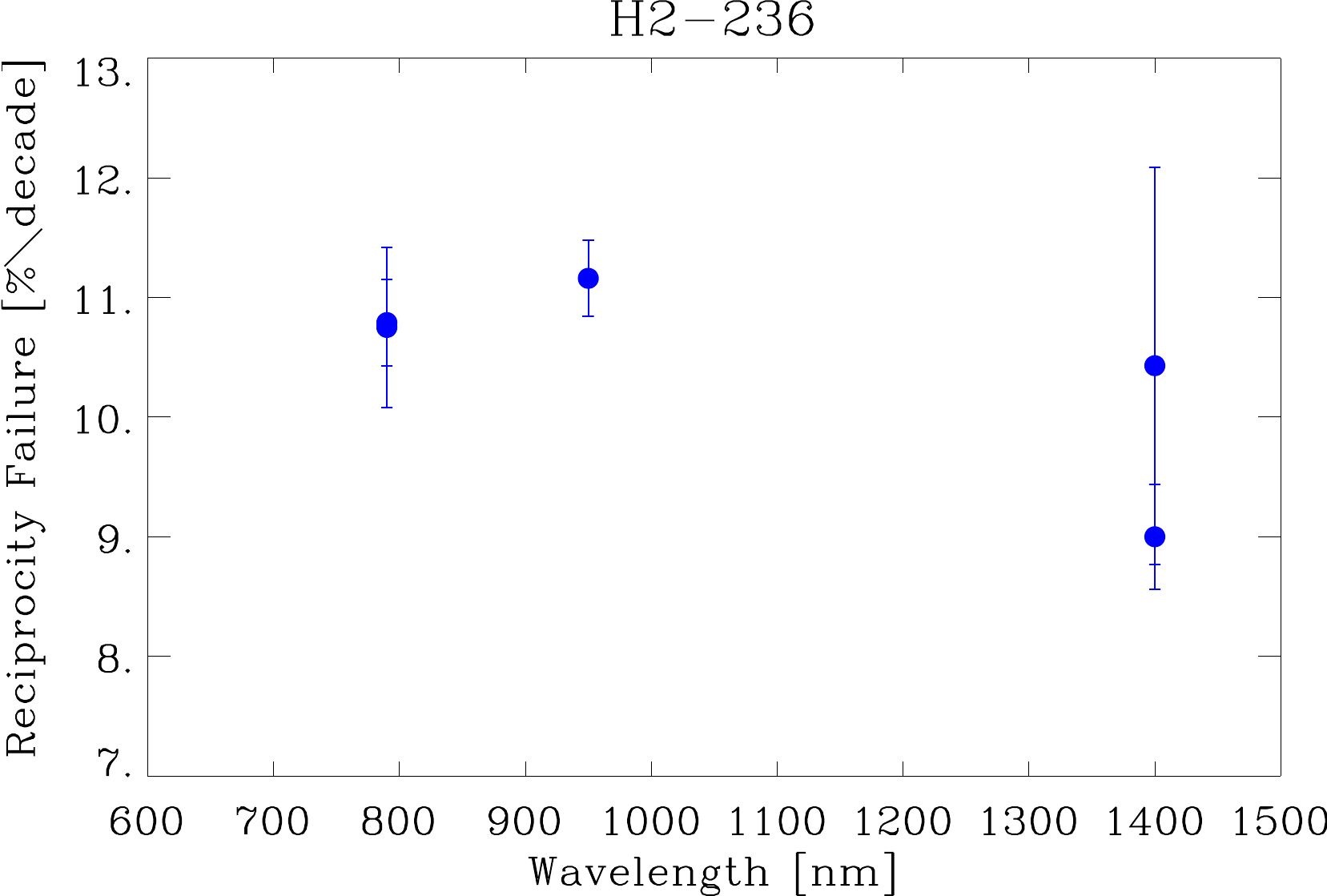}
\end{tabular}
\end{center}
\end{minipage}
\caption[rec_failure_wave]{Reciprocity failure as a function of wavelength for H2-102 (left panel) and H2-236 (right panel).}
\label{fig:rec_failure_wave}

\end{center}
\end{figure}

A test station for the measurement of reciprocity failure on NIR detectors was built. It achieved a sensitivity of 0.3\,\%/decade. 
So far four 1.7\,$\mu$m HgCdTe arrays that had been fabricated for SNAP/JDEM were tested. In two of the devices (H2-102 \& H2-142) only a small amount of reciprocity failure was observed. Two other devices  (H2-236 \& H2-238) show significant reciprocity failure and strong spatial variation. Measurements were performed at five wavelengths between 700\,nm and 1400\,nm. No indication for wavelength  dependency was found (see Fig.\,\ref{fig:rec_failure_wave}). This contrasts the reported behavior of the NICMOS detectors on HST. 
Fabrication of JDEM/SNAP devices is based on WFC3 detector development. This is reflected in measurements on the final candidate detectors for WFC3 which show very similar results as H2-102 and H2-142\cite{Hill_garching}. The WFC3 team reports reciprocity failure ranging from 0.3\,\%/decade to in 
to 0.97\,\%/decade for three detectors. As they point out, this is significantly smaller than the effect in the 2.5\,$\mu$m HgCdTe NICMOS detectors on HST (6\,\%/decade).
NICMOS detector material was grown using the liquid phase epitaxy technique while molecular beam epitaxy was used for the growth of material for the WFC3 and for the SNAP R\&D detectors.

%-------------
%-------------

%1.1?m) effect seen for NICMOS detectors on HST, which are 
%an earlier generation of HgCdTe detectors with a different long 
%wavelength cutoff (2.5 ?m).

The two detectors H2-236 and H2-238 clearly stand out. We investigated possible correlations of the observed spacial structure in the reciprocity failure of detector H2-236 with other measured detector characteristics such as QE, dark current, conversion gain, read noise and persistence (see Table\,\ref{table:summary}). No correlation was observed. However, it is intriguing that the detector with highest persistence (H2-236) also shows the strongest reciprocity failure. A detailed investigation of a possible correlation between persistence and reciprocity failure continues to be part of our ongoing test program.

\begin{table}
\label{tab:summary}
\begin{center}       
\begin{tabular}{|l|c|c|c|c|} 
\hline
\rule[-1ex]{0pt}{3.5ex} NIR Device & H2-102 & H2-142 & H2-236 & H2-238   \\
%\rule[-1ex]{0pt}{3.5ex}  Name & Position 1 & Position 2 &
%Wavelength [nm]&  (FWHM [nm])  \\
\hline
\hline
\rule[-1ex]{0pt}{3.5ex}  Reciprocity Failure [\%/decade] & 	&	&	&	\\
\rule[-1ex]{0pt}{3.5ex}  \hspace{.2cm} 700\,nm	- strip & 0.26$\,\pm\,$0.04	&	&	&	\\
\rule[-1ex]{0pt}{3.5ex}  \hspace{.2cm} 790\,nm	- strip & 0.27$\,\pm\,$0.03	&0.30$\,\pm\,$0.03	& 10.79$\,\pm\,$0.36	&\\
\rule[-1ex]{0pt}{3.5ex}  \hspace{.2cm} 790\,nm	- full & & 0.45$\,\pm\,$0.12		& 10.75$\,\pm\,$0.67	&\\
\rule[-1ex]{0pt}{3.5ex}  \hspace{.2cm} 880\,nm	- strip & 0.28$\,\pm\,$0.06	&	&	&	\\
\rule[-1ex]{0pt}{3.5ex}  \hspace{.2cm} 950\,nm	- strip & 0.21$\,\pm\,$0.04	&		&	11.16$\,\pm\,$0.32	&4.27$\,\pm\,$0.42\\
\rule[-1ex]{0pt}{3.5ex}  \hspace{.2cm} 950\,nm	- full & 	&		&		&4.21$\,\pm\,$0.71\\
\rule[-1ex]{0pt}{3.5ex}  \hspace{.2cm} 1400\,nm - strip& 0.30$\,\pm\,$0.05&0.24$\,\pm\,$0.04		&9.00$\,\pm\,$0.44	&	\\
\rule[-1ex]{0pt}{3.5ex}  \hspace{.2cm} 1400\,nm - full& & 	&10.43$\,\pm\,$1.66	&	\\
\hline
\rule[-1ex]{0pt}{3.5ex}  Pedestal material & Molybdenum & SiC & SiC & SiC  \\
\rule[-1ex]{0pt}{3.5ex}  CDS read-noise	 [$e^-$]		& 25.0	& 24.8	& 25.7	& 22.3	\\
\rule[-1ex]{0pt}{3.5ex}  Dark current	[$e^-/s/pix.$]	& 0.02	& 0.03	& 0.03	& 0.05	\\
\rule[-1ex]{0pt}{3.5ex}  Persistence	[\%]			&	& $<$0.02	& 0.2	& 0.26	\\
\hline 

\end{tabular}
\end{center}
\caption[filters]
{\label{table:summary}Comparison of the magnitude of reciprocity failure and typical characteristics for the tested  HgCdTe detectors. All measurements were performed at a temperature of 140\,K. Median values are given for read noise and dark current.}
\end{table} 

All measurements reported here have been conducted with the detectors at 140\,K. Preliminary measurements at lower temperatures indicate a reduction in the amount of reciprocity failure. More tests are under way to investigate a possible correlation between detector temperature and reciprocity failure. Such a correlation would suggest charge traps in the material causing the observed effect.

\acknowledgments     %>>>> equivalent to \section*{ACKNOWLEDGMENTS}       

We thank Curtis Weaverdyck for invaluable help with setting up the experiment.
This work was supported by the Director, Office of Science, of the
U.S. Department of Energy under Contract No. DE-FG02-08ER41566.
% Much of the
%%original set-up has been influenced by a similar system set up for
%measurements of CCDs at Lawrence Berkeley Laboratory. For the measurement of
%illumination flatness and the dramatic improvement we greatly
%benefited from the help of one of our summer students, Cesar Palma.

%%%%%%%%%%%%%%%%%%%%%%%%%%%%%%%%%%%%%%%%%%%%%%%%%%%%%%%%%%%%%
%%%%% References %%%%%

\bibliography{report}   %>>>> bibliography data in report.bib
\bibliographystyle{spiebib}   %>>>> makes bibtex use spiebib.bst

\end{document}